\documentclass[amsmath, amssymb, twocolumn]{revtex4-2}

\usepackage{graphicx}
\usepackage{hyperref}
\usepackage{upgreek}

\newcommand{\slmsuite}{\href{http://github.com/slmsuite/slmsuite}{\texttt{slmsuite}}}
\newcommand{\order}{$\mathcal{O}$}

\begin{document}
\title{Full-volume aberration-space holography}
\author{Ian~Christen}
\email{ichr@mit.edu}
\author{Christopher~Panuski} 
\author{Thomas~Propson}
\author{Dirk~Englund} 
\email{englund@mit.edu}
\address{
Research Laboratory of Electronics, Massachusetts Institute of Technology, Cambridge, MA 02139, USA
}
\date{\today}
\begin{abstract}
Simultaneous, diffraction-limited control of multiple optical beams is crucial for applications ranging from lithography to optogenetics, deep tissue imaging, and tweezer-based manipulation of cells, particles, or atoms.
Despite the desire to address wider fields of view, deeper volumes, and increasingly-disordered media, spatially-varying aberrations currently restrict parallelized steering to a limited ``isoplanatic" region over which the point spread function is invariant.
Here, we overcome this limitation by combining individual propagation kernels accounting for site-specific aberrations into a single spatial light modulator (SLM) hologram. 
This ``aberration-space holography" unlocks precise, parallel holographic shaping over the SLM's entire Nyquist-limited volume, enabling us to realize full-field, anisoplanatic aberration compensation for the first time.
By simultaneously correcting 50 isoplanatic patches with 8 principal aberration modes, we demonstrate a full-field 
optical tweezer array with ${8\times}$ larger field of view than the best isoplanatic correction.
Extending to 3D, we increase the volume of a multiphoton volumetric display by ${12\times}$.
These performance enhancements are immediately accessible to a diverse range of applications through our open-source software implementation, which combines aberration-space holography with automated experimental feedback, wavefront calibration, and alignment.
\end{abstract}

\maketitle

\section{Introduction}
Aberrations are inherent to the practical optical systems driving technical progress and scientific discovery.
The open goal of efficient, programmable optical routing to overcome these imperfections is therefore central to imaging and microscopy~\cite{bertolotti2022imaging, miller2020cellular, yoon2020deep}, optogenetic photostimulation~\cite{adesnik2021probing, russell2022all, zhou2023deep}, \textit{in-vivo} micromanipulation~\cite{bunea2019strategies, favre2019optical}, targeted chemical assembly~\cite{zhang2024genetically}, photolithography and laser fabrication~\cite{somers2024physics, salter2019adaptive, panuski2022full}, atom array construction and control~\cite{kaufman2021quantum, manetsch2024tweezer, christen_2022_integrated}, atmospheric turbulence compensation~\cite{davies2012adaptive, guo2024direct}, holographic displays~\cite{smalley2018photophoretic}, and beyond.
In each case, ideal aberration compensation would enable diffraction-limited readout or addressing over the entire accessible volume.
While this goal has traditionally been approached passively through the incorporation of additional static components (compound lenses or compensated objectives, for example)~\cite{botcherby2012aberration}, the number of required elements---alongside the associated cost, complexity, and optical loss---grows with numerical aperture, field of view size, and desired correction order. 

\begin{figure*}[ht]
\centering
\includegraphics[scale=1.3]{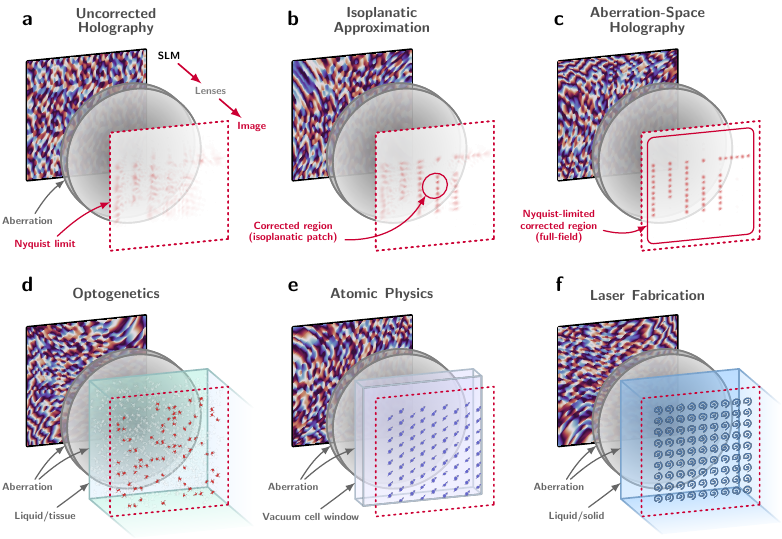}
\caption[Anisoplanatic Aberration Compensation and Applications]{
\textbf{Anisoplanatic aberration compensation and applications.}
\textbf{a,} 
Fourier holography with an uncorrected SLM phase mask fails to reproduce target patterns/images due to inherent, spatially-dependent aberrations.
\textbf{b,} 
Whereas previous efforts use a global phase correction to optimize image quality within a given neighborhood (isoplanatic patch)
\textbf{c,} 
our method enables near-diffraction-limited holography throughout the SLM's entire addressable field of view (i.e., Nyquist volume)---which includes 3D depth along the direction of propagation (not pictured)---by combining individual corrections for each targeted site.
This aberration-space holography enables full-volume steering for applications ranging from 
\textbf{d,} 
targeted imaging and excitation in disordered media, 
\textbf{e,} 
single-atom tweezing through vacuum cell cover glass, and 
\textbf{f,} 
selective material modification for laser machining in solids or resins. SLM masks are simulated and phase wrapped at $2\pi$ for illustration.
} \label{fig_applications}
\end{figure*}

Alternatively, the emerging field of programmable optics holds significant promise as a scalable solution for wide-field, dynamic correction in complex environments~\cite{hampson2021adaptive}. Adaptive optics (AO) implementations, for example, enable near-diffraction-limited astronomy from ground-based telescopes by physically shaping few-degree-of-freedom macroscopic deformable mirrors to invert atmospheric phase distortions~\cite{ghez2008measuring}. 
Recently, the advent of spatial light modulators (SLMs) with \order(millions) of control degrees of freedom has extended these capabilities to wavefront shaping at the wavelength-scale~\cite{vellekoop2007focusing}. Similar to deformable mirrors, phase-based SLMs enable low-loss steering for power-sensitive applications such as deep tissue imaging with strict damage thresholds~\cite{qin2022deep}, tweezer-based optical trapping with limited input power~\cite{barredo2018synthetic, manetsch2024tweezer}, or computing accelerators seeking to improve energy efficiency~\cite{bernstein2023single, mcmahon2023physics}. 
The principal benefit of SLMs, therefore, is extending phase control to orders of magnitude more degrees of freedom. 
Typical SLMs operate in the pupil-conjugate configuration shown in  Fig.\ref{fig_applications}a and use ``Fourier holography" with indirect feedback for image formation. 
In this case, holography reshapes a nearfield wavefront into a desired farfield pattern using phase retrieval techniques such as the Gerchberg-Saxton (GS) algorithm, which incorporates a propagation kernel approximating the ideal Fourier transform properties of free-space propagation and lensed focusing~\cite{di2007computer, nogrette2014single, kim2019large}.

These techniques should enable high-fidelity optical control anywhere within the ``Nyquist volume": the positions in three-dimensional space where the spatial frequencies of the source phase mask are resolvable by the pixels of the SLM. 
However, to date, systems are limited by ubiquitous spatially-dependent refractive index perturbations---resulting from, for example, imaging through vacuum windows~\cite{manetsch2024tweezer}, deep tissues~\cite{ji_2010_neighborhood, papadopoulos2017scattering, papadopoulos2020dynamic, may_2021_neighborhood}, or turbulent fluids~\cite{guo2024direct}---an effect that is exacerbated as numerical aperture increases~\cite{jesacher2010parallel}.
These so-called ``anisoplanatic" aberrations prevent simultaneous operation over the complete field of view~\cite{thurman2008correction}.

Instead, adaptive optics has thus far been, at best, optimized for correction at a single ``isoplanatic patch" (IP), outside of which anisoplanatic aberrations dominate.
In astronomy, the \order(10 arcsecond) field of view accessible around isoplanatic guidestar correction limits the speed at which wider images can be stitched together from serially-corrected IPs~\cite{guyon2018extreme}.
While state-of-the art wavefront sensors can measure single-IP aberration in real time~\cite{guo2024direct}, parallel correction over multiple IPs remains an open goal due to the limitations of both tomographic and multiconjugate AO (reduced image quality from averaged corrections and system complexity from multiple wavefront sensors/correctors, respectively)~\cite{hampson2021adaptive, ragazzoni2000adaptive, rigaut2018multiconjugate}. 
Similar issues are amplified in microscopy, where \order(1-100 $\upmu$m) IPs are common in complex media such as biological samples~\cite{wang2015direct, liu2019direct}. 
Guidestar descanning~\cite{wang2014rapid}, scattering matrix reconstruction~\cite{popoff2010measuring, badon2020distortion}, and other serial techniques~\cite{qin2020adaptive, blochet2023fast} for multiple-IP corrections are limited by sample dynamics and prevent real-time observation~\cite{zhang2023adaptive}. 
Alternatively, pupil segmentation enables spatial multiplexing at the expense of numerical aperture~\cite{park2017large, may2021simultaneous}. 
Recent computational~\cite{haim2024image} and multi-spectral techniques~\cite{balondrade2024multi} offer avenues to improve parallelization; however, widefield correction over multiple IPs remains fundamentally limited.

Here, we consider the effects of anisoplanatism for the reciprocal application: light projection to points across the Nyquist volume.
Sample applications (Fig.~\ref{fig_applications}d-f) of such optical focus arrays include neutral atom quantum computing~\cite{barredo2018synthetic, manetsch2024tweezer}, holographic displays~\cite{smalley2018photophoretic}, high-speed 3D laser printing~\cite{zhang2024high}, and multifocal nonlinear microscopy~\cite{wu2021speed, zhang2019kilohertz}.
In each case, anisoplanatic aberration limits the achievable scale, speed, and parallelism.
Extending subtractive wavefront correction from imaging to projection (Fig.~\ref{fig_applications}b) enables aberration compensation near a single calibration point~\cite{papadopoulos2017scattering} but prevents parallel use of the SLM's full Nyquist volume.
As with imaging, serial correction is a common work-around albeit with limited speed due to an SLM's temporal response ~\cite{wang2020aberration}.

\begin{figure*}[ht]
\centering
\includegraphics[scale=1.3]{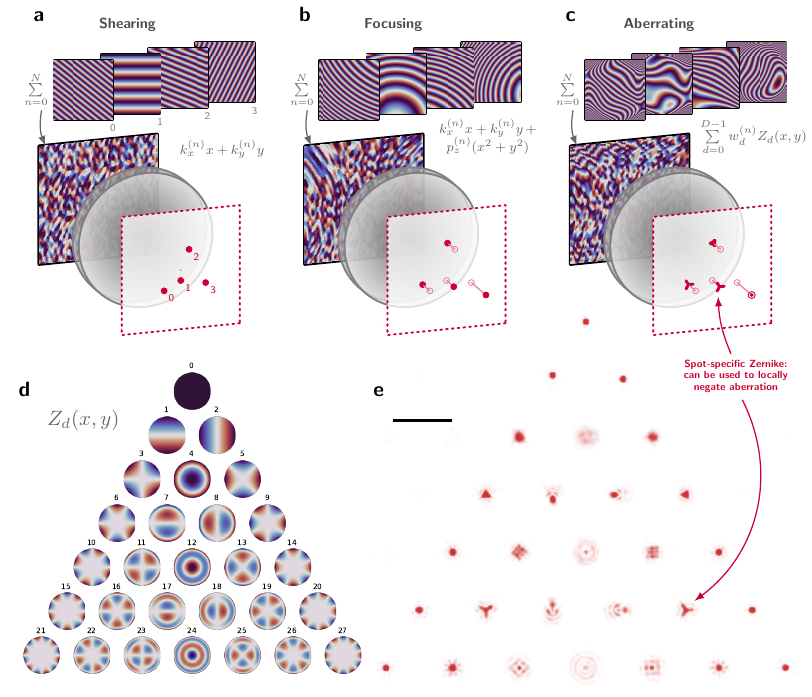}
\caption[Parallel Beamshaping across Aberration Space]{
\textbf{Parallel beamshaping across aberration space.}
\textbf{a,} 
The Fourier transform uses linear ``shearing kernels" 
to steer light at a fixed focal depth.
\textbf{b,} 
Fresnel holography extends this concept to three-dimensional space through ``focusing kernels" that include a parabolic term proportional to focal power $p_z^{(n)}$.
\textbf{c,} 
To fully generalize this technique, we introduce ``Zernike kernels" targeting points in a $D$-dimensional \textit{aberration space} formed by basis functions $Z_d$. Shaping individual spots according to a 
\textbf{d,} 
Zernike pyramid, for example, yields 
\textbf{e,} 
an experimental array of corresponding spots sourced from a single hologram. The scalebar corresponds to 100~$\upmu$m at the camera. Alternatively, tailoring each spot to negate Zernike coefficients $w_d$ corrects anisoplanatic aberration.
} \label{fig_zernike}
\end{figure*}

Full-field, parallel beamforming has yet to be realized by existing techniques. 
To address this open goal, we propose and demonstrate ``aberration-space holography" to compensate for anisoplanatic deviations of a system's propagation kernel from the ideal Fourier transform. As illustrated in Fig.~\ref{fig_applications}c, we show that generating and merging individual propagation kernels---expanded in an optimized basis set of aberrating functions (Fig.~\ref{fig_zernike})---for each target point enables complete and customizable control anywhere within the Nyquist volume. 
This technique unlocks the full volume of holographic shaping that SLMs \textit{should} be capable of.

\section{Aberration-Space Holography}\label{sec_zernike}
Aberration-space holography generalizes the angle-space ($k$-space) steering characteristics of the Fourier holography kernel to a complete basis set of aberrating functions.
In Fourier holography, the phase of a nearfield pattern $f(x,y)$ that 
approximates a desired farfield amplitude distribution $|G(k_x,k_y)|$ in $k$-space is often computed through a series of iterative discrete Fourier transforms (DFTs) $\mathcal{F}_\text{DFT}$:
\begin{align*}
    f(x,y) = \mathcal{F}_\text{DFT}^{-1}\{F\} \propto &\sum_{k_x,~k_y} K^*(k_x,k_y,x,y) \cdot F(k_x,k_y),
    \\
    F(k_x,k_y)  = \mathcal{F}_\text{DFT}\{f\} \propto &\sum_{x,~y} K(k_x,k_y,x,y) \cdot f(x,y)
\end{align*}
Here, $K(k_x,k_y,x,y) = \exp(-i[k_xx + k_yy])$ is the linear ``shearing kernel", or precisely the SLM phase that steers $f(x,y)$ to the farfield angle $(k_x,k_y)$. 
Summing these kernels provides the lateral steering illustrated in Fig.~\ref{fig_zernike}a; \textit{how} to achieve the optimum sum is solved by the particular algorithm. 
In GS, for example, the nearfield and farfield during each iteration are corrected to match the source and target amplitudes, respectively, until $|F(k_x,k_y)|$ converges to $|G(k_x,k_y)|$.

The insight of so-called ``compressive sensing" spot holography~\cite{pozzi2018fast} is that $|G(k_x,k_y)|$ is typically sparse: only a small subset of $N$ farfield pixels are targeted, especially when considering computation on megapixel-scale SLMs (which can yield gigapixel-scale padded farfields to achieve the desired angular resolution). 
The remaining points computed via DFT are generally zeroed or ignored.
In this case, the DFT can be equivalently written as a compressed sum $\mathcal{F}_\text{c}$ of kernels $K_n(x,y)$---each corresponding to a farfield point of interest $(k_x^{(n)}, k_y^{(n)})$:
\begin{align*}
    f(x,y) = \mathcal{F}_c^{-1}\{F\} &\propto \sum_{n}^{N} K_n^*(x,y) \cdot F_n, \\
    F_n  = \mathcal{F}_c\{f\} &\propto \sum_{x,~y} K_n(x,y) \cdot f(x,y).
\end{align*}
Spot holography then distills to choosing the $N$ complex farfield weights $F_n$ which, similar to DFT-based holography, can be solved with iterative algorithms (e.g. GS). 
As a first immediate benefit, this technique frees target spot locations from the 2D DFT grid, alleviating the need for zero-padding (alongside the associated memory and computing requirements) to achieve the desired positioning precision.

More importantly, having now abstracted propagation from the Fourier transform's shearing kernel, $K_n(x,y)$ can be tailored to account for additional propagation phenomena. For example, adding a parabolic phase forms a ``focusing kernel" (Fig.~\ref{fig_zernike}b) enabling 3D spot arrays (i.e., Fresnel holography)~\cite{di2007computer, chew2024ultraprecise}.
The phase arguments of the kernels demonstrated to date are specific cases of a general modal expansion of the pupil wavefront phase. 
For example, the summation $\sum_dw_d^{(n)}Z_d(x,y)$ of Zernike polynomials for a circular pupil accounts for both in-plane and depth steering since $x \propto Z_2 = Z_1^1$, 
$y \propto Z_1 = Z_1^{-1}$, and $\delta z \propto x^2 + y^2 \propto Z_4 = Z_2^{0}$. 
Note that we index each polynomial $Z_d = Z_n^l$ with radial (azimuthal) order $n$ ($l$) by a singleton ANSI index $d$ as illustrated in Fig.~\ref{fig_zernike}d.

Individually, these generalized kernels enable free-form structured light at any singular target point.
Diffraction-limited spots can therefore be achieved by tailoring each kernel to negate local aberration. 
While previous work has focused on this implementation of isoplanatic correction~\cite{neil2000closed, booth2002adaptive},
our technique instead combines kernels for \textit{all} $N$ points of interest into a single hologram for full-field control (Fig.~\ref{fig_zernike}c). The experimental Zernike pyramid in Fig.~\ref{fig_zernike}e, for example, is generated with a single hologram incorporating nearly two waves ($\pm5$ radians peak-valley) of the respective Zernike kernel at each point in the pyramid.

Instead of reshaping power in two- or three- dimensional space, our technique for holography can be viewed as reshaping power within a $D$-dimensional \textit{aberration space}. 
This concept extends beyond shearing and focusing kernels or analytic correction for depth-dependent spherical aberration ($Z_{12} = Z_4^0$)~\cite{jesacher2010parallel}, instead enabling complete anisoplanatic control across 2D fields of view and 3D volumes as illustrated in Secs.~\ref{sec_field} and \ref{sec_volume}, respectively.
In the case of correction, a set of kernels constructed from a basis of $D$ Zernike polynomials offers fully parallelized performance approaching that of a single $D^\text{th}$ order correction.

\section{A Full-Field 2D Tweezer Array}\label{sec_field}
\begin{figure*}[tp]
\centering
\includegraphics[width=\textwidth]{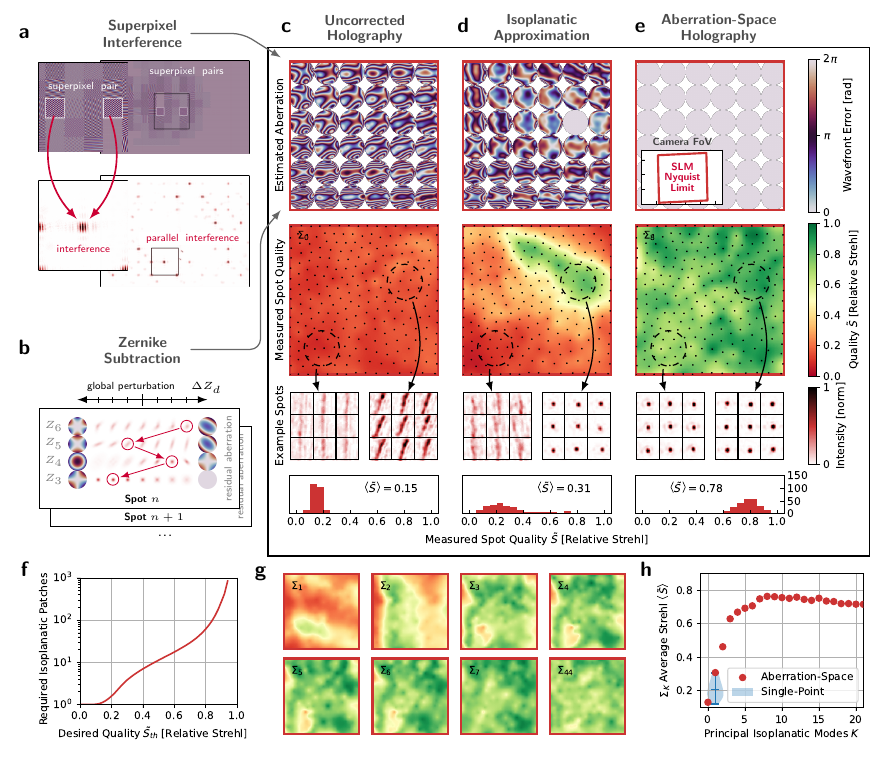}
\caption[A Full-Field Tweezer Array via Anisoplanatic Wavefront Measurement and Correction]{
\textbf{A full-field tweezer array via anisoplanatic aberration measurement (a-c) and correction (d-h).}
To project through strong anisoplantic aberration, we developed two parallelized techniques for wavefront measurement over the field of view:
\textbf{a,} farfield interference (bottom) between ``superpixel" pairs diffracted to common points via applied phase gratings (top), which provides their relative phase by fitting imaged fringe patterns; and 
\textbf{b,} Zernike subtraction, where we sweep each Zernike order $Z_d$ with a global perturbation $\Delta Z_d$ to locally measure and iteratively subtract aberration (thereby minimizing spot size as illustrated by the path of optimization) at each target site. 
\textbf{c,}
The resulting spatially-dependent wavefront error can be visualized across the SLM's Nyquist-limited field of view (red outlines throughout), as illustrated here by a sample 6$\times$6 grid of residual aberration maps. We remove linear shearing terms defining position, leaving only higher-order, local aberrations.
\textbf{d,}
Existing techniques subtract site-specific wavefront error at a point of interest (site $(3,5)$ of the illustrated grid), leaving residual aberration outside this isoplanatic patch.
The illustrated correction maximizes IP size; other correction points with larger aberration gradients yield smaller IPs.
\textbf{e,} 
Alternatively, our method corrects aberration across the SLM's entire Nyquist-limited field of view.
\textbf{c-e,}
The interpolated spot quality (measured by relative Strehl ratio $\tilde{S}$ at each black target point) for each correction case illustrates the $8\times$ diffraction-limited ($\tilde{S}>0.8$) field-of-view enhancement achieved by aberration-space holography.
Example spot images (21 pixels/50 $\upmu$m square on the camera) from clusters outside (left) and inside (right) the isoplanatic neighborhood of correction further evidence our ability to extend near-diffraction-limited performance beyond the isoplanatic correction region.
The most aberrated spots (left three of the six grids) are shown with $4\times$ color scale for visibility.
$\tilde{S}$ histograms depict overall quality statistics for each case.
\textbf{h,}
For comparison, we estimate a lower bound for the number of isoplanatic patches needed to fully correct the Nyquist-limited field of view to above a threshold Strehl $\tilde{S}_{th}$.
\textbf{g,}
The anisoplanatic correction in \textbf{e} is optimized by projecting the measured aberration matrix onto the $K$ most significant principal isoplanatic modes, which produces the correction $\Sigma_K$.
\textbf{f,}
We find that $K=8$ maximizes averaged Strehl $\langle \tilde{S}\rangle$. $\Sigma_1$, the principal isoplanatic correction, outperforms any single-point isoplanatic correction (blue violin plot distribution).
} \label{fig_calibration}
\end{figure*}

To demonstrate this principle, we generated 2D optical tweezer arrays with full-field correction to $D=44^\text{th}$ order, yielding the results in Fig.~\ref{fig_calibration}. 
The setup is intentionally-aberrated (using strongly warped acrylic; see Methods\ref{meth:setups}) and designed to image the entire Nyquist-limited farfield of the SLM. 
Since compensated projection requires accurate aberration characterization, we first developed two parallel techniques~(Fig~\ref{fig_calibration}a-b) to measure anisoplanatic wavefront error.

The first extends camera-based interferometric readout~\cite{vcivzmar2010situ} from one to many farfield measurement points. This method extracts the relative phase between two patches of SLM pixels, or ``superpixels", by analyzing their interference pattern at a desired point. 
Our parallel extension simultaneously deflects power from multiple superpixel pairs to measurement points across the entire field of view~(Fig.~\ref{fig_calibration}a), reducing measurement time by the factor of parallelization: \order(100$\times$) in our case. 
This method also measures the amplitude distribution at the SLM as required for high-performance computer generated holography.

\begin{figure*}[ht]
\centering
\includegraphics[width=\textwidth]{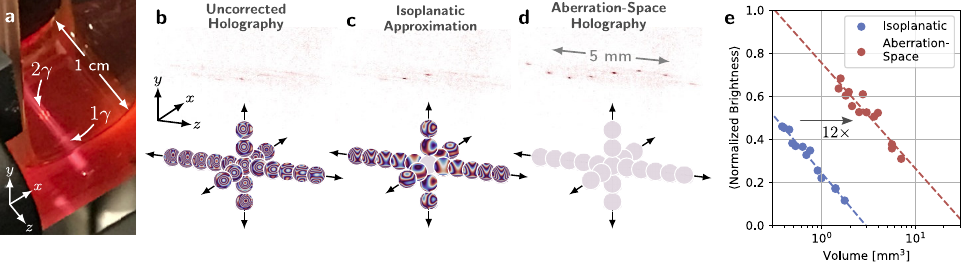}
\caption{
\textbf{An aberration-corrected volumetric display.}
\textbf{a,}
Our display comprises a solution of quantum dots in 1-cm-wide cuvette, producing fluorescence at tightly focused spots (here, in the form of a three-dimensional cube) via two-photon excitation ($2\gamma$).
Short-pass wavelength filters block one-photon scatter ($1\gamma$)  in subsequent measurements.
To illustrate the depth enhancement enabled by aberration-space holography, $2\gamma$ fluorescence from a projected helix of spots is shown under the same three conditions as Fig.~\ref{fig_calibration}c-e:
\textbf{b,}
without aberration correction,
\textbf{c,}
with isoplanatic correction, and
\textbf{d,}
using our technique for full-volume correction.
Each panel contains a view looking slightly off-axis from $\hat x$.
Background scattering and zero-order diffraction florescence are subtracted computationally for visual clarity.
Below each florescence image, we also show the estimated residual aberration---extrapolated from Zernike subtraction measurement---along cardinal axes.
Quadratic defocus terms are removed to highlight point-specific deviations from the ideal Fresnel kernel.
\textbf{e,}
To quantify the field of view enhancement, we evaluate average spot brightness (normalized to peak brightness) for a cube centered about the SLM's zeroth order (as depicted in \textbf{a}) and include the depth improvement measured from the helix, revealing a $12\times$ increase in addressable volume.
} \label{fig_volume}
\end{figure*}

Further parallelization can be achieved with our second method which directly optimizes spot quality at each farfield target point by iteratively subtracting aberration~(Fig.~\ref{fig_calibration}b). 
After generating a baseline hologram targeting the desired points, we superimpose and sweep a global Zernike order to determine its contribution at each point based on a spot quality figure of merit.
Aberration-space holography then allows us to subtract the measured anisoplanatic correction from each spot.
By iterating with global Zernike orders up to $D$, we converge to an aberration-space spot array that negates local aberration to within measurement accuracy.
Notably, this Zernike subtraction method is applicable to non-imaging-based indirect feedback metrics such as atomic trap depth or two-photon florescence intensity.

Following calibration, we use weighted GS algorithms~\cite{kim2019large} to project full-field tweezer arrays under three wavefront correction conditions (Fig.~\ref{fig_calibration}c-e): (1) uncorrected, (2) isoplanatic, and (3) anisoplanatic via aberration-space holography. 
In each case, we estimate residual wavefront error, analyze per-site optical focus quality $\tilde{S}$ (measured as the relative Strehl ratio normalized to its peak value across all spot arrays; see Methods\ref{meth:strehl}), and present $\tilde{S}$ statistics across the array.

As expected due to our significantly aberrated setup, uncorrected holography (Fig.~\ref{fig_calibration}c) produces distorted spots throughout the field of view, yielding $\langle \tilde{S}\rangle = .15$ and a maximum $\tilde{S}$ of .27. 
The optimum isoplanatic correction (Fig.~\ref{fig_calibration}d)---selected to maximize the number of points with $\tilde{S}>.8$---significantly improves performance within a local region, enabling near-unity $\tilde{S}$ near the correction point. 
However, full-field performance remains limited: $\sim$20\% of the Nyquist-limited field of view is moderately compensated ($\tilde{S} > .5$) and even less, about 2\%, achieves the traditional Mar\'{e}chal criterion for diffraction-limited performance ($\tilde{S} > .8$). 
Other isoplanatic corrections yield significantly smaller corrected regions, especially around the edges of the field of view where the aberration gradient is steepest.
To estimate the minimum number of IPs needed to correct the entire Nyquist-limited field of view to a desired threshold $\tilde{S}_{th}$, we calculated and integrated the isoplanatic patch density (Fig.~\ref{fig_calibration}h, Methods\ref{meth:area}).
More than 50 patches are required to satisfy $\tilde{S}_{th} > .8$ and serially correcting each may be infeasible depending on experimental time constraints.

In contrast, aberration-space holography (Fig.~\ref{fig_calibration}e) enables full-field, parallelized correction with $\langle \tilde{S}\rangle = .78$, a minimum $\tilde{S}$ of .47, and 43\% of the SLM's field of view corrected to $\tilde{S} > .8$.
To optimize the anisoplanatic correction achievable with aberration-space holography, we first projected our $D = 44^\text{th}$-order calibration (measured via Zernike subtraction) onto $K$ principal isoplanatic modes using singular value decomposition~\cite{badon2020distortion} as illustrated in Fig.~\ref{fig_calibration}g, h (Methods\ref{meth:svd}).
These principal isoplanatic modes are a set of basis vectors (Fig.~\ref{fig_principal}), derived by rotating the Zernike basis and ordered by normalized singular values $\tilde{\sigma}_i=\sigma_i/\sum_i\sigma_i$ describing their relative significance. 
The associated Shannon entropy $H=\sum_i\tilde{\sigma}_i\log_2\tilde{\sigma}_i\approx 4.22$ indicates that a distilled correction $\Sigma_K$ with $K=5$ isoplanatic modes captures the essential information content of each isoplanatic patch; however, increasing correction order to $\Sigma_8$ (Fig.~\ref{fig_pyramid}) produces the best averaged Strehl $\langle\tilde{S}\rangle$ across the field of view (a $\sim 20\%$ improvement from $\Sigma_5$). 
Less-significant principal modes are interpreted to contain residual noise from the calibration process (Fig.~\ref{fig_violin}).
This decomposition also reveals $\Sigma_1$, the principal isoplanatic correction, which outperforms all single-point isoplanatic corrections in terms of $\langle\tilde{S}\rangle$.

\section{A Corrected Volumetric Display}\label{sec_volume}
Beyond addressing planar wavefront error, aberration-space holography also mitigates aberration throughout a volume.
As a proof-of-concept demonstration of parallelized holography across adverse volumes, we examine the efficiency of two-photon excitation within a cuvette-bound solution of cadmium selenide zinc sulfide quantum dots  (Fig.~\ref{fig_volume}a, Methods\ref{meth:setups}).
This setup is highly sensitive to depth-dependent spherical aberrations associated with projecting into the cuvette~\cite{ren2014three, diel2020tutorial, jesacher2010parallel}, an ``anisovolumism" generalizing focal plane anisoplanatism.
Under two-photon excitation, a hologram of pulsed near-infrared ($\sim$800~nm) light excites bright and isolated spots of orange ($\sim$610~nm) quantum dot fluorescence at the hologram's focal points.
This nonlinear excitation---with fluorescence depending on the square of optical power---allows us to evaluate spot quality based on emission brightness.

We thus calibrate aberration throughout the medium using $D=27^\text{th}$-order Zernike subtraction to maximize spot brightness (Fig.~\ref{fig_volume}b-d).
This is again done in parallel using a camera (facing $\hat{x}$) to image and calibrate arrays of spots in $\hat{y}$-$\hat{z}$ planes.
Given any three-dimensional point within the calibration volume, we fit and interpolate the resulting calibration point aberration maps to determine the $D$-dimensional aberration vector that optimizes focus.

This process enables us to dynamically project arbitrary three-dimensional patterns with optimum brightness.
Using a high-speed SLM (Santec SLM-210 with  $\tau$$<$10~ms rise time), our system acts as a video-rate, aberration-corrected volumetric display with \order{($10^8$)} addressable voxels ($\sim 3 \times 3 \times 34$~$\upmu$m$^3$ spots
across a $\sim 3.75\times3.75\times10$~mm$^3$ volume).
Here, the display volume is bounded laterally by the Nyquist limit and in depth by the cuvette dimensions.
Larger volumes or tighter spots are possible by increasing or decreasing, respectively, the effective focal length $f$ of the system, which acts as the constant of proportionality $k_x \approx x/f$ between the nearfield and farfield.

To quantify the field-of-view enhancement enabled by aberration-space holography, we measure the average spot brightness of a variable-volume, 8-point cube projection centered about the SLM's zeroth diffraction order to estimate lateral display area. 
Combined with the measured display depth enhancement, this reveals a logarithmic relationship between brightness and enclosed display volume for both isoplanatic and aberration-space corrections (Fig.~\ref{fig_volume}e).
Since anisoplanatism-independent lateral effects---including SLM diffraction efficiency and chromatic dispersion, which can be compensated with dispersive lenses~\cite{bouchal_2014_achromatic, hu_2016_achromaticaod, guo_2019_achromaticdmd}---are common to both correction curves, we conclude that anisoplanatic correction enables a $\sim12\times$ increase in addressable volume.

\section{Discussion}
Our first-ever demonstration of parallelized, full-field aberration correction---which enabled in our experiments more than an order of magnitude greater utilization of the imaging field or volume---illustrates the immediate applicability of aberration-space holography to a diverse range of technologies currently bottlenecked by imperfect physical optics. 
For example, \order(10$^4$ site) scaling limits in state-of-the-art neutral atom arrays~\cite{manetsch2024tweezer}, $>$\order(100 mm) focal length designs for near-eye holographic displays~\cite{schiffers2024holochrome}, and \order($10^8$~voxel/s) printing speeds in multi-foci two-photon printers~\cite{zhang2024high} can all be overcome with the presented techniques. 
Imaging applications requiring confocal scanning can also be parallelized and accelerated by scanning an aberration-compensated spot array instead of serially interrogating isoplanatic patches.
For rapid integration into these applications, our open-source software \slmsuite{} combines a user-friendly, computationally-efficient implementation of the associated algorithms with experimental feedback, hardware calibration, and automated alignment. 

From a theoretical perspective, aberration-space holography extends a core imaging principle to projection: efficient use of \textit{all} degrees of freedom. 
Due to aberration, state-of-the-art imagers often contain more sensing degrees of freedom (camera pixels) than the number of spots resolvable through the associated optical system. 
Light field imagers therefore use microlens arrays or other tiled optics to extract additional information (depth measurements, for example) from previously unused degrees of freedom~\cite{wu2022integrated, guo2024direct}. 
Analogously, aberration typically prevents simultaneous diffraction-limited projection to $M$ farfield voxels with an $M$-pixel SLM. By combining individual shaping kernels for each targeted point into a single hologram, our approach programmatically reconfigures SLM pixels for more efficient widefield control. 

In the ideal case without aberration, propagation from an SLM with $M$ degrees of freedom is represented by a Fourier transform. In this case, the basis of $M$ orthogonal shearing kernels in the nearfield is perfectly translated to a grid of $M$ orthogonal diffraction-limited spots in the farfield.
With anisoplanatic aberration, the farfield basis is distorted: 
the spots are spread outside their respective diffraction-limited gridsites with overlapping magnitudes even though the complex fields remain orthogonal.
While it is mathematically feasible to untangle the knot of each distorted spot in the farfield---finding the delicate balance of complex nearfield parameters to produce a sharp farfield hologram~\cite{badon2020distortion, yu2024complex}---it is inefficient to do so directly for large basis sizes.
Instead, our method untangles this knot in the nearfield by replacing the linear shearing kernels with nonlinear nearfield kernels appropriate to produce a basis of diffraction-limited spots in the farfield.
In this way, we can recover the $M$ distinct spots that the SLM's degrees of freedom are intrinsically capable of generating.

Compared to recent machine learning-based aberration compensation techniques~\cite{gopakumar2024full}, our semi-analytic approach enables tunable, deterministic, per-site shaping that opens new avenues for ultra-high-resolution 3D holography beyond inter-plane crosstalk limits~\cite{makey2019breaking}. 
Conversely, we anticipate that combining our spatially-resolved aberration maps with deep learning strategies will aid in understanding the location and structure of wavefront error~\cite{tippie2010multiple}, simplifying multi-conjugate adaptive optics in astronomy~\cite{johnston1994analysis} and beyond.
These combined characterization and compensation strategies drive towards practical, aberration-free optical systems to enable the next generation of discovery.

\section{Acknowledgments}
I.C. acknowledges support from the National Defense Science and Engineering Graduate Fellowship Program and the National Science Foundation (NSF) award DMR-1747426.
C.P. acknowledges support from the Hertz Foundation Elizabeth and Stephen Fantone Family Fellowship.
T.P. acknowledges support from the NSF Graduate Research Fellowship Program and the MIT Jacobs Presidential Fellowship.
The authors thank all \slmsuite{} contributors, Santec for providing a SLM-210 demo unit, as well as Sixian You and Li-Yu Yu for providing femtosecond laser light.

\section{Data availability}

Data are available from the corresponding authors upon reasonable request.

\section{Code availability}
Our open-source implementation of aberration-space holography---alongside other functionality for SLM control, SLM calibration, and data analysis---is available at \url{https://github.com/slmsuite/slmsuite}.

\section{Author Contributions}
I.C. conceived the aberration-space holography idea and acquired the data.
I.C., C.P., and T.P. developed the codebase \slmsuite{}.
I.C. and C.P. developed the experiments, led data analysis, and wrote the manuscript with input from all the authors.
D.E. supervised the project and motivated analysis techniques including SVD.

\appendix
\renewcommand{\thesection}{}

\begin{figure*}[ht]
\centering
\includegraphics[width=\textwidth]{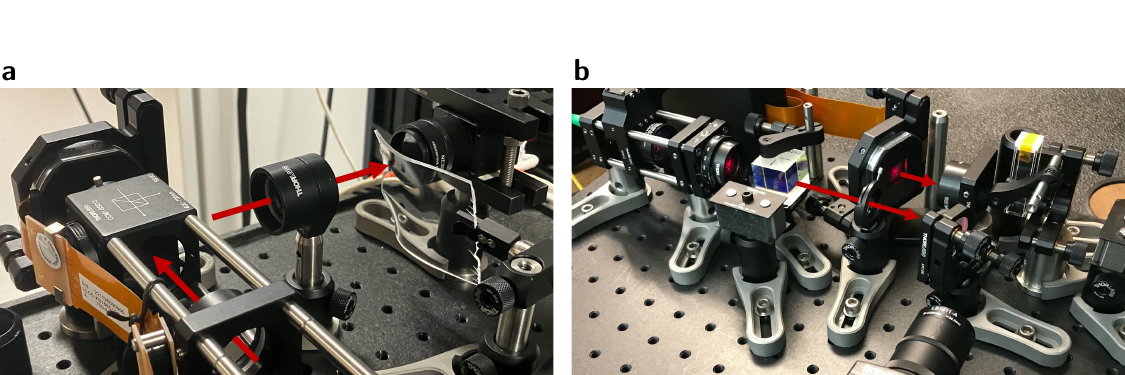}
\caption{
\textbf{Experimental setups.}
\textbf{a,}
2D setup with warped acrylic aberration (right side).
\textbf{b,}
3D setup, here shown with an empty cuvette (right side).
} \label{fig_setup}
\end{figure*}

\begin{figure*}[ht]
\centering
\includegraphics[width=.75\textwidth]{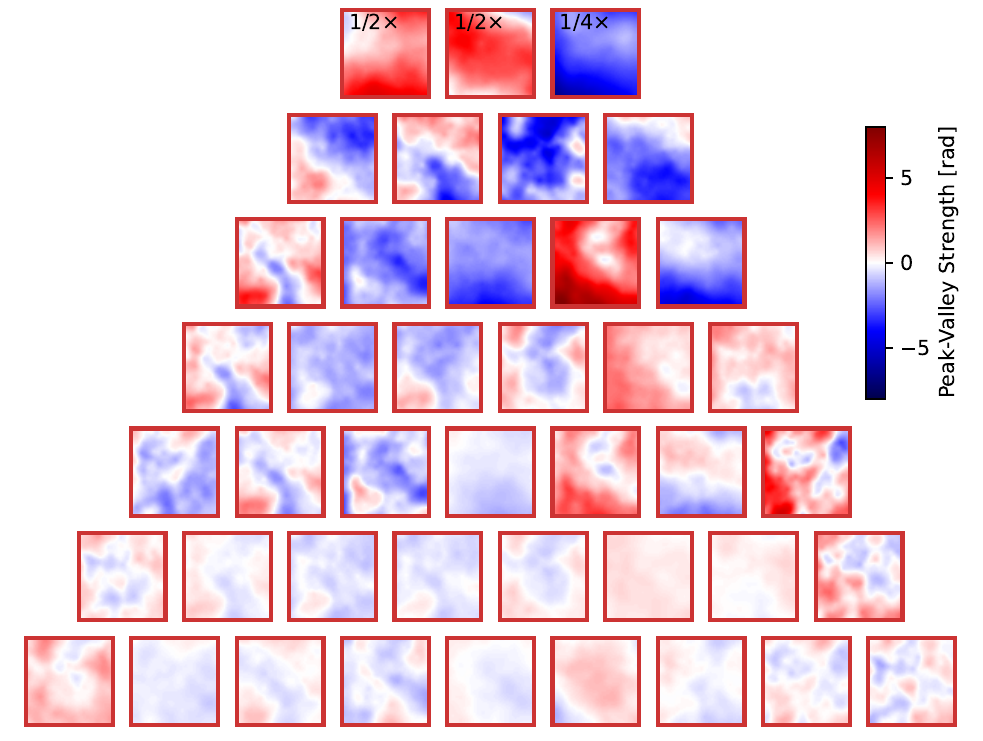}
\caption{
\textbf{Measured anisoplanatic aberration.}
The coefficients of each Zernike term for $\Sigma_8$ (Fig.~\ref{fig_calibration}e) arranged in the standard pyramid (see Fig.~\ref{fig_zernike}).
The lower-order terms are strongest: indeed, on this plot, the first three terms ($Z_3, Z_4, Z_5$) are scaled (as labeled) to fit within the plot's color scale.
Patterns between terms are revealed, for instance as a similarity throughout columns that suggests correction from a common principal isoplanatic mode.
} \label{fig_pyramid}
\end{figure*}

\begin{figure*}[ht]
\centering
\includegraphics[width=\textwidth]{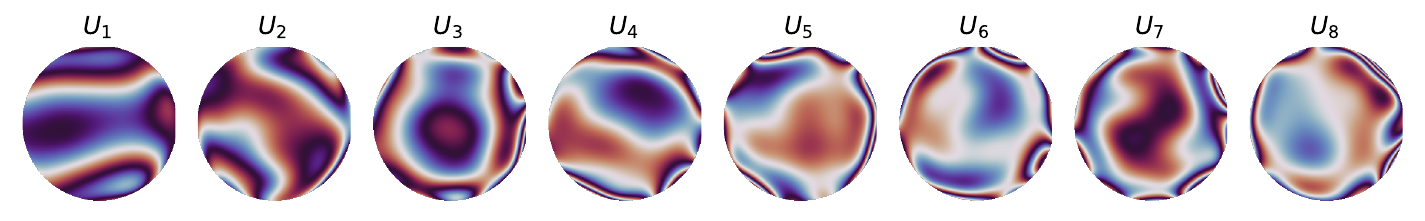}
\caption{
\textbf{Principal isoplanatic modes.}
Plots of the first eight principal modes derived from SVD analysis.
$U_i$ is the $i^\text{th}$ column of the $U$ basis rotation.
$U_1$ is dominated by low-order terms, while higher-order principal corrections are increasingly dependent on higher-order terms.
} \label{fig_principal}
\end{figure*}

\begin{figure*}[ht]
\centering
\includegraphics[width=.9\textwidth]{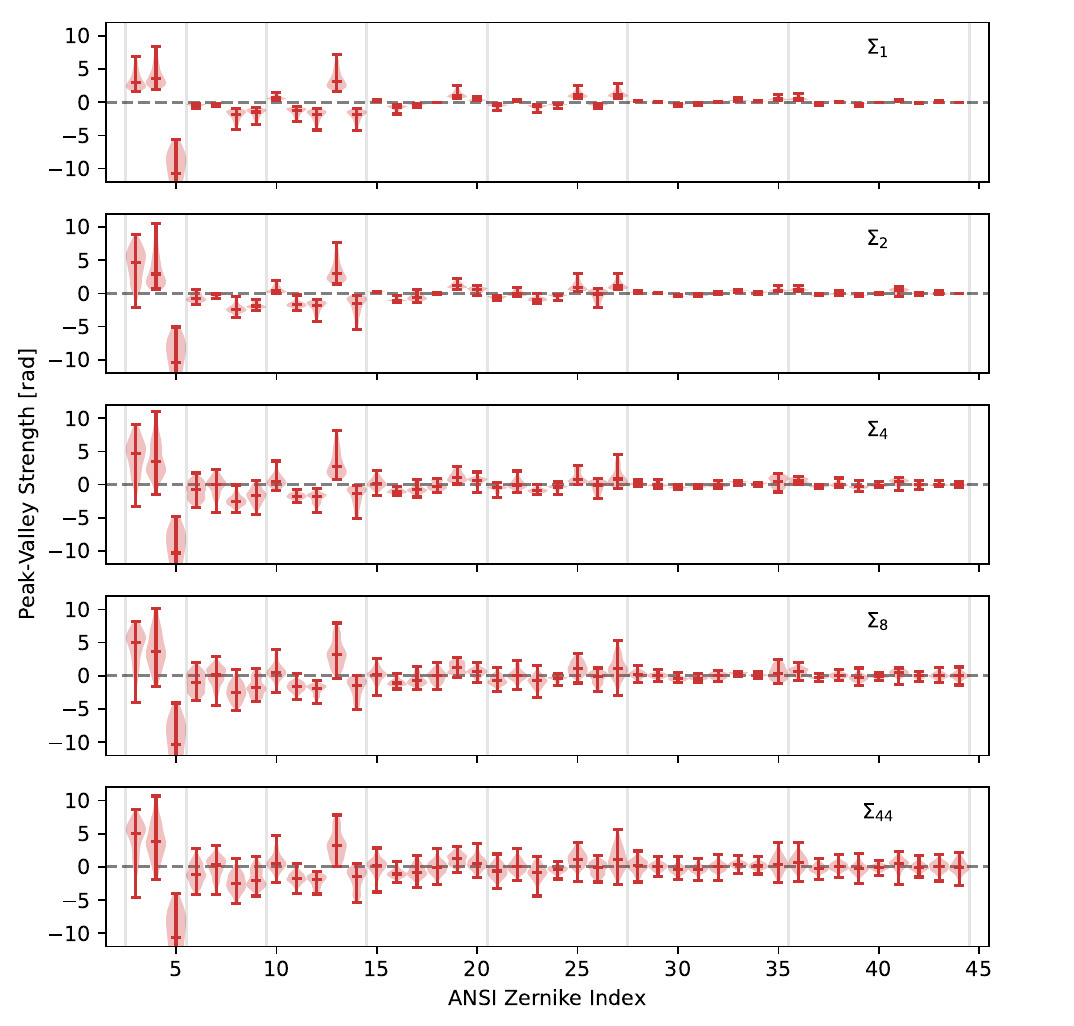}
\caption{
\textbf{Increasing anisoplanatic correction.}
The data corresponding to $\Sigma_8$ is a processed version of the raw data in Fig.~\ref{fig_pyramid}, where each violin consists of the histogram and box and whisker statistics of the range of the anisoplanatic correction across the Nyquist limit.
For $\Sigma_1$, only one principal mode is present, scaled in strength across the plane.
Thus, each violin as the same shape, but is scaled according to index.
From $\Sigma_1$ to $\Sigma_8$, more principal modes are added, acting as a correction largely to the higher-order terms upon the strong low-order base correction.
$\Sigma_{44}$ shows stronger variance in the higher orders, which is interpreted as excess noise.
In all cases, $Z_5$ extends down to a correction of nearly -30 radians (not shown here to focus on the fine structure of the higher order terms).
} \label{fig_violin}
\end{figure*}

\begin{figure}[ht]
\centering
\includegraphics[width=.5\textwidth]{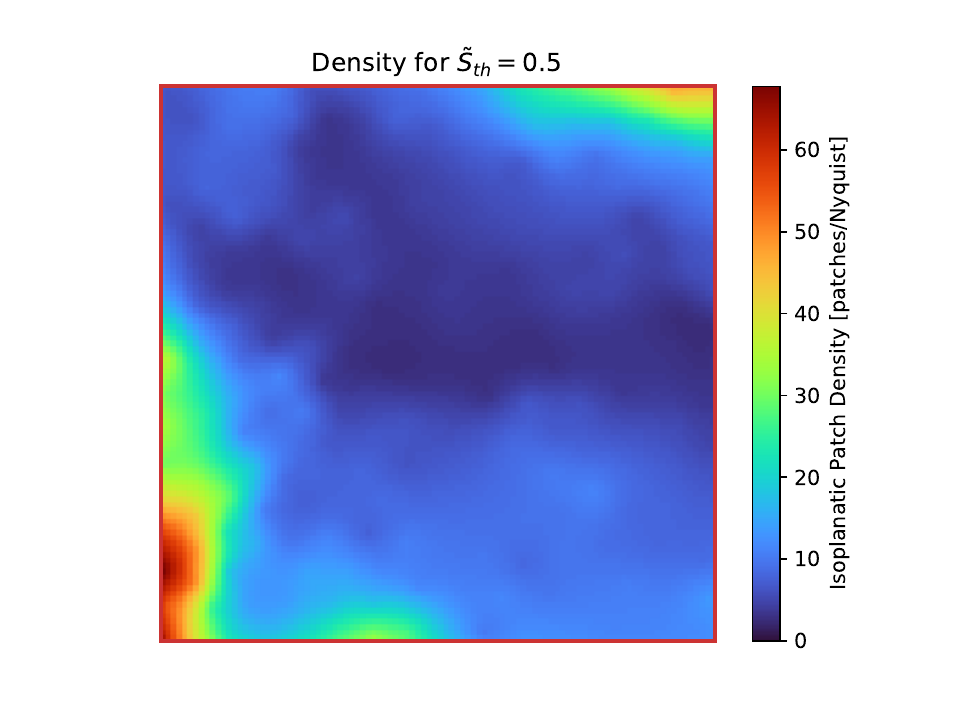}
\caption{
\textbf{Isoplanatic patch density.}
A map of the estimate of patch density for $\tilde{S}_{th} = .5$ across the Nyquist limit.
This is scaled such that the integrated value across the space is equivalent to the result displayed in Fig.~\ref{fig_calibration}f.
The large area in the upper middle co-indices with the largest isoplanatic patch as chosen in Fig.~\ref{fig_calibration}d, as expected.
} \label{fig_density}
\end{figure}
\section*{Methods}
\subsection{Computational Scaling}\label{meth:gpu}

The majority of computational holography's complexity is generating nearfield-to-farfield transformation kernels along each $k$-vector. 
For 2D Fourier holography, the linear shearing kernels are efficiently implemented by the discrete Fourier transform (DFT).
Working beyond the 2D case requires more complicated techniques operating on now-nonlinear kernels: overlap integrals using the kernels (near to far) and summation of many such kernels (far to near).

For general nonlinear kernels, DFTs cannot be used and the kernel data must be directly evaluated to realize the transforms.
Caching this data is tractable for \order{(100)} spots, and we implement a CPU-/GPU- compatible version of compressed holography using vectorized operations on these kernels in \texttt{numpy}/\texttt{cupy}/\texttt{pytorch}.
However, the memory cost of caching kernel data is prohibitive for large numbers of spots.
For instance, when generating a hologram with \order(10$^4$) spots and 32-bit precision using a megapixel SLM, a large \order(40~GB) cache is required: a task reserved for the most expensive GPUs.

Fortunately, more commonplace GPUs (e.g., the NVIDIA GeForce GTX 1070 Ti used for these experiments) provide an immediate solution to this problem.
Instead of precomputing and caching kernels, we can dynamically compute them during each propagation.
Here, each of the GPU's parallel threads locally compute the desired Zernike basis for a specific pixel of the megapixel nearfield.
That is, each distinct Zernike polynomial 
\begin{align*}
Z_d(x,y) = \sum_{n,m} a_{nm}^{(d)} x^ny^m
\end{align*}
is evaluated at pixel $(x,y)$ using only global information regarding the $a_{nm}^{(d)}$ coefficients of the monomial terms $x^ny^m$.
Then, this Zernike basis is used to construct the kernels for each spot via a $N\times D$ matrix $w^{(n)}_d$ corresponding to the $D$-dimensional vectors describing the $N$ spot positions in aberration-space.
Different bases (such as principal isoplanatic modes) can be used to minimize $D$ or accomplish other goals.
Compared to the time cost of loading data onto the GPU, locally computing the basis in register memory is fast and amortized over many farfield spots (each pixel-specific Zernike basis is used to compute each spot's propagation kernels).
Changing the aberration-space coordinates of spots also remains agile, as these are operations on the relatively small data encoded in $w^{(n)}_d$ alongside the $N$ complex weights $F_n$ describing the farfield.
In this way, dynamic video-rate aberration-space holography can be achieved.

The results presented in this work focus on compressed spot holography, but are not limited to sparse schemes.
As a first iteration, these techniques are immediately applicable to the correction of local aberration in image fanout~\cite{bernstein2023single}.
Beyond this, our techniques could be extended to dense image holography, where each kernel pointing to a vector in aberration space could be augmented with a local miniature DFT-based chirp-$Z$ transformation expanding dense 2D holography within the purview of the isoplanatic patch of each compressed spot.

Additionally, while we focus on WGS algorithms for farfield spots, some precision applications---
such as refining extremely uniform top hats or images---are better optimized by conjugation gradient descent methods.
Our cached implementation of aberration-space holography supports forward- and backward- propagation of the gradients of the complex Fourier weights via \texttt{pytorch}.

\subsection{Experimental Setups}\label{meth:setups}
In our baseline setup (Fig.~\ref{fig_setup}a), fiber-coupled 633~nm light (HeNe) is collimated and expanded to the size of the SLM (Santec SLM-200).
A 50:50 beamsplitter separates the incident and reflected paths, with the reflected path focused (Thorlabs 150~mm achromat) to our camera (Huateng Vision HT-SUA2000M-T1V-C, 20.0~MP xor AlliedVision Alvium 1800 U-158, 1.6~MP).
The 20 megapixel (MP) camera resolves the full farfield of the SLM, whereas the lower-resolution camera captures localized datasets (Zernike pyramid, some wavefront calibration). An optical attenuator (OD3) blocks noise from room light, while still allowing the bright laser through. For the full-field correction demonstration (Fig.~\ref{fig_calibration}), we image through $\approx$3~mm-thick, thermally-warped acrylic to produce stronger anisoplanatism in our low-NA setup.

The volumetric display (Fig.~\ref{fig_setup}b) consists of a fiber-coupled femtosecond laser (Spectra-Physics Mai Tai, $\sim$100~fs, $\sim$800~nm) collimated in free-space to the size of the SLM (Thorlabs 50~mm achromat) and polarized 
 with a polarizing beamsplitter (Thorlabs PBS252).
After reflecting off the SLM (Santec SLM-210) at $\sim 15^\circ$ incidence, the light is focused (Thorlabs 35~mm achromat) into a cuvette (Alpha Nanotech UV quartz; 1~cm pathlength) containing cadmium selenide zinc sulfide quantum dots suspended in a liquid toluene solution (NN-Labs HECZ600, 2~mg/mL).
Two cameras image the setup: 
a camera perpendicular to the direction of excitation (Thorlabs DCC1545M, 1.3~MP) and an off-axis 8.2~mm endoscope tube camera (Vividia, color 2.0~MP). The tube camera images are used for perspective (Fig.~\ref{fig_volume}b-d), while the perpendicular camera images are used for calibration.
Short pass filters remove the 1-photon scattering from both imagers.

\subsection{Relative Strehl Metrology}\label{meth:strehl}

The Strehl ratio, a measure of maximum spot intensity relative to the ideal point spread function, is a standard metric to judge imaging quality.
To estimate Strehl from camera images $I$, we first numerically calculate a discrete metric proportional to Strehl:
\begin{align*}
S' = {\max_{x\in X, y\in Y} I}\left/{\sum_{x\in X, y\in Y} I}\right.,
\end{align*}
i.e, the maximum image intensity within boxes around each spot position (defined by positions $X$, $Y$) normalized to the box's integrated energy.
Our experiments use $21 \times 21$ pixels integration windows, nearly an order of magnitude larger than the diffraction-limited spot waist (approximately 3 pixels).
We then normalize $S'$ to its maximum value across all experiments (except for those in Methods\ref{meth:area}), yielding relative Strehl $\tilde{S}$.

\subsection{Isoplanatic Patch Density}\label{meth:area}

To judge the challenge of anisoplanatic correction, we quantify the difficulty of the alternative: serial interrogation of $P$ isoplanatic patches within which $\tilde{S}$ remains above a desired threshold $\tilde{S}_{th}$.
Rather than estimating $P$ by tiling single-point isoplanatic corrections (e.g., Fig.~\ref{fig_calibration}d) to fully cover the Nyquist-limited field of view, we instead integrate the local patch density derived from these results across the full field.

Specifically, we measure relative Strehl $\tilde{S}$ for $J=256$ single-point isoplanatic corrections that are evenly spaced across the full field of view. For each correction, we linearly interpolate $\tilde{S}$ to produce an approximately continuous estimate of spot quality and estimate the local isoplanatic patch density by inverting the measured area $A^{(j)}_{\tilde{S} > \tilde{S}_{th}}$ above the threshold $\tilde{S}_{th}$. 
Smaller areas $A^{(j)}_{\tilde{S} > \tilde{S}_{th}}$ imply a stronger anisoplanatic gradient.
We can further convert this to a lower bound for the total number of required anisoplanatic patches via the equation:
\begin{align*}
P(\tilde{S}_{th}) = 
\frac{1}{J}\sum_{j=0}^J \frac{A_{tot}}{A^{(j)}_{\tilde{S} > \tilde{S}_{th}}}.
\end{align*}
Most corrections measure a maximum Strehl $\tilde{S}_{max} < 1$, so to avoid division by zero area when considering $\tilde{S}_{th} > \tilde{S}_{max}$, we instead normalize each correction to $\tilde{S}_{max}$. This normalization enlarges $A^{(j)}_{\tilde{S} > \tilde{S}_{th}}$ and thereby underestimates local patch density and $P$.

The results in Fig.~\ref{fig_calibration}f can be understood in the context of two specific cases.
For sufficiently small $\tilde{S}_{th}$, $A^{(j)}_{\tilde{S} > \tilde{S}_{th}}=A_{tot}$ and $P=1$.
Alternatively, for evenly distributed isoplanatic patches throughout the plane without overlap, $A^{(j)}_{\tilde{S} > \tilde{S}_{th}} = {A_{tot}}/{J}$ and $P(\tilde{S}_{th}) = J$ isoplanatic patches as expected.

\subsection{Anisoplanatic Singular Value Decomposition}\label{meth:svd}
We adopt a singular value decomposition (SVD) approach to rotate the $D$-dimensional Zernike basis into a basis of $D$ principal isoplanatic modes~\cite{badon2020distortion}.
Here, the Zernike matrix of our $N$-point full-field correction $W = w_d^{(n)}$ (size $N\times D$) is expressed as
\begin{align*}
    W = U \Sigma V^\dagger,
\end{align*}
where the coefficients of the diagonal matrix $\Sigma$ (size $D\times D$) are singular values
corresponding to the relative importance of principal isoplanatic modes, i.e. the singular vectors, of our anisoplanatic system (Fig.~\ref{fig_principal}).
$U$ (size $N\times D$) and $V$ (size $D\times D$) perform the basis rotation between the Zernike basis (size $D$) and the principal isoplanatic modes (size $D$).

Restricting our correction to the $K$ most significant principal modes (by zeroing diagonal $\Sigma$ terms beyond the first $K$) yields the correction $\Sigma_K$ that maximizes information density to $K^\text{th}$ order (Fig.~\ref{fig_violin}).
Beyond revealing the highest-performance (de-noised) correction $\Sigma_{8}$ in Fig.~\ref{fig_calibration}f, this analysis also reveals the information-theoretic ``best average" isoplanatic correction $\Sigma_1$. This correction outperforms any of the isoplanatic corrections which subtract the aberration at a specific point.

\bibliography{bib}

\end{document}